# Measurement-Based Modeling and Analysis of UAV Air-Ground Channels at 1 and 4 GHz

Zhuangzhuang Cui, César Briso-Rodríguez, *Senior Member, IEEE*, Ke Guan, *Senior Member, IEEE*, César Calvo-Ramírez, Bo Ai, *Senior Member, IEEE*, and Zhangdui Zhong, *Senior Member, IEEE*

*Abstract*—In the design of unmanned aerial vehicle (UAV) wireless communications, a better understanding of propagation characteristics and an accurate channel model are required. Measurements and comprehensive analysis for the UAV-based air-ground (AG) propagation channel in the vertical dimension are presented in this letter. Based on the measurement data at 1 and 4 GHz, the large-scale and small-scale channel parameters are extracted in the line-of-sight (LOS) and nonLOS case, respectively. The altitude-dependent path loss model is proposed herein. Furthermore, shadow fading and fast fading are statistically analyzed for comprehensively describing the fading behavior. Our results will be useful in the modeling of AG channels and the performance analysis for UAV-enabled wireless communication systems.

## I. Introduction

THANKS to their prominent features of high mobility and flexible deployment, unmanned aerial vehicles (UAVs), will find many promising uses in the future fifth generation (5G) and beyond wireless communications [1]. With continuous cost reduction and device miniaturization, small UAVs (typically with weight not exceeding 25 kg and with a maximum altitude of 120 m) are now more easily accessible to the public. Hence, numerous new applications in the civilian and commercial domains have emerged, with typical examples including weather monitoring, forest fire detection, traffic control, cargo transport, emergency search and rescue, communication relay, and others [2]. Among these applications, UAV can be operated in different trajectories such as horizontal and vertical flight, which bring many challenges for air-ground (AG) propagation channel modeling. First, the dimension of modeling has evolved into three-dimension (3-D) from the traditional two-dimension, which means that it is critical to pay more attention to the variable altitudes of UAV and thereby build up an altitude-dependent channel model. Second, the frequency band for UAV communications can be various such as L-band, C-band, and millimeter-wave band. Thus, the impact of the alternative frequencies on propagation characteristics is supposed to be considered in channel modeling. Since the uniform regulation of frequency band has not yet been formed, International Telecommunication Union Radiocommunication Sector (ITU-R) recommends that part of L-band can be used for the control and nonpayload communication (CNPC) for UAV and C-band is used primarily for payload transmission [3].

In related works, measurement campaigns were conducted for different scenarios such as urban and near-urban at L-band and C-band in [4], where tapped delay line (TDL) models were presented. However, since the measurements were carried out at a single altitude, it is insufficient for setting up an altitude-dependent model. Authors in [5] and [6] carried out some measurements with vertical flights of UAV at 2.585 and 4.3 GHz, respectively. However, the impact of frequency was neglected. In [7], the measurements for drone-to-ground channels were conducted in the line-of-sight (LOS) and nonLOS (NLOS) cases in different altitudes at 900, 1800, and 5 GHz, in which log-distance path loss model was used to predict the fading behavior. In our previous work, we conducted some wideband measurements and ray-tracing simulations for the channel characterization at 3.9 GHz, but lacked a specific altitude-dependent channel model and with a single frequency. Thus, the topic of this letter is to analyze and model the impact of the frequency and the altitude of UAV.

The remainder of this letter is organized as follows. In Section II, we introduce the measurement setup, including the scenario description and the measurement system and data acquisition. Section III presents the measurement results and corresponding models. Besides, the shadowing fading and fast fading are statistically analyzed. The fading depth and the theoretical distribution are discussed for further describing the fading behavior. Conclusions are drawn in Section IV.

## II. Measurement Setup

### A. Scenario Description

As shown in Fig. 1(a), the measurement campaign was conducted on the campus. In the measurement, the UAV integrated the transmitter (Tx) equipment flies in an open field and the ground station (GS) serving as the receiver (Rx) is placed on the roof of a building. The UAV ascends from an open field area at a distance 350 m from the GS.

Manuscript received July 5, 2019; accepted July 19, 2019. Date of publication July 23, 2019; date of current version September 4, 2019. This work was supported in part by the National Key R&D Program of China under Grant 2016YFB1200102-04, in part by the NSFC under Grant 61771036, Grant 61911530260, Grant U1834210, and Grant 61725101, and in part by the State Key Laboratory of Rail Traffic Control and Safety (Contract RCS2019ZZ005), Beijing Jiaotong University. *(Corresponding author: Ke Guan.)*

Z. Cui, K. Guan, B. Ai, and Z. Zhong are with the State Key Lab of Rail Traffic Control and Safety, Beijing Jiaotong University, Beijing 100044, China (e-mail: 17111020@bjtu.edu.cn; kguan@bjtu.edu.cn; boai@bjtu.edu.cn; zhdzhong@bjtu.edu.cn).

C. Briso-Rodríguez and C. Calvo-Ramírez are with the ETSIS Telecomunications, Universidad Politécnica de Madrid, 28031 Madrid, Spain (e-mail: cesar.briso@upm.es; cesarrasec4@gmail.com).

Digital Object Identifier 10.1109/LAWP.2019.2930547

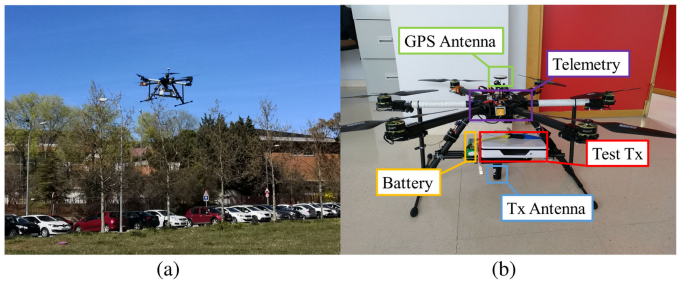

Fig. 1. UAV. (a) Photograph of the test environment with the UAV in the flight. (b) Detailed composition of hexacopter UAV integrated Tx equipment.

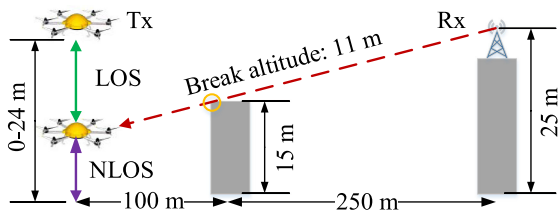

Fig. 2. Detailed view of essential parameters in the measurement scenario.

### B. Measurement System and Data Acquisition

The system consists of a small and lightweight airborne transmission module and a GS. Fig. 1(b) depicts the DJI N3 aircraft fully integrated with the test Tx unit together with the battery, telemetry unit, global positioning system (GPS) antenna, and a Tx antenna. The UAV belongs to the type of "small" with a weight of 4.7 kg and can be operated in low altitudes. The battery and the telemetry give support on the power and control for UAV flight, respectively. The GPS module can provide accurate height. For the transmitted signal, a continuous wave is used. There are two channels in the Tx equipment: one is 0.4–1.1 GHz, the other is 3.4–5 GHz. In our measurements, we focus on the two narrowband measurements (1, 4 GHz) that correspond to L-band and C-band. The transmitted power at both frequencies is 30 dBm, and the noise floor of Rx is 120 dBm. Thus, the link margin is 150 dB, providing enough power to do measurements in NLOS case. The parameters of measurement are given in Table I.

TABLE I
MEASUREMENT SETUP

| Transceiver | Freq. | Height | Antenna | Polarization |
|---|---|---|---|---|
| Tx/UAV | 1/4 GHz | 0–24 m | RM-WHF | Vertical |
| Rx/Analyzer | 1/4 GHz | 25 m | $\lambda/4$ monopole | Vertical |

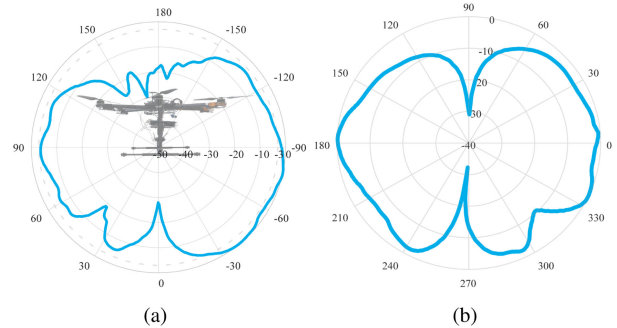

Fig. 3. E-plane radiation pattern at 4 GHz. (a) Tx antenna. (b) Rx antenna.

The detailed setup of measurement and the size of the main buildings are drawn in Fig. 2. The altitude of UAV goes from 0–24 m ($\pm 1$ m), and the height of GS is set to 25 m, which corresponds to the parameters in the 3GPP channel model. In the 3GPP TR 38.901 [8], the path loss model in the urban macro (UMa) scenario is applicable to 25 m for base station and 1.5–22.5 m for the user equipment (UE).

Because of the blockage of building with a height of 15 m, the link is NLOS when the altitude of UAV is from 0 to 11 m according to the geometric relationship. When higher than 11 m, the LOS is no longer obstructed. Thus, the setup of flight and the choice of location provide the possibility to model LOS and NLOS cases of AG channels.

In addition, the antenna of Tx is the Rugged Surface mount-wideband high frequency (RM-WHF) antenna, which is approximately omnidirectional and with 3 dBi gain. The Rx incorporates a signal analyzer of type MS2830A and a $\lambda/4$ monopole antenna with 2.15 dBi gain. Since the ground plane is not very good, the gain of $\lambda/4$ monopole antenna is smaller than the theoretical value. It also should be noticed that both antennas in the Tx and Rx are vertically polarized. The E-plane radiation patterns of Rx and Tx at 4 GHz are plotted in Fig. 3, and the H-plane radiation patterns of both ends are approximately omnidirectional. In measurements, the antenna elevation angles used for measurements are $-30°(330°)$ to $30°$ and $-120°$ to $-60°$ for Rx and Tx, respectively. Moreover, the horizontal distance is 350 m and the height of UAV is from 0 to 24 m, and the elevation angle $\theta = \arctan(\frac{h_R - h_T}{D})$ between Tx and Rx is very small with a range of $0.16°$–$4.09°$. Such small range of $\theta$ can ensure the antenna pattern used for measurements have a good alignment.

We carried out multiple measurements to ensure the size of measured data is large enough and the reliability of data is high enough. Five independent round-trip measurements in the vertical dimension were conducted for each frequency, and we selected a set of data obtained with the most stable flight of UAV. The data are averaged by the round-trip data for reducing the deviation. Moreover, in the storage of signal analyzer of Rx, the received signal was stored 62.5 per meter in distance, and each stored data is an average of 20 collected data for each altitude of UAV. Thus, the total number of measured data for two frequencies has reached 120 000 ($24 \times 2 \times 20 \times 62.5 \times 2$). With a large number of measured data, the channel parameters can be characterized with negligible deviation.

### III. PROPAGATION MODELING AND MEASUREMENT RESULTS

#### A. Path Loss

The commonly used path loss model in AG channels is the free-space (FS) model given by [9]

$$PL_{\text{FSPL}} = 32.45 + 20\log_{10}(d) + 20\log_{10}(f) \quad (1)$$

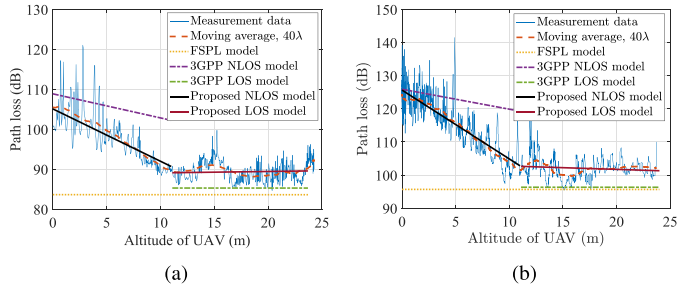

Fig. 4. Path loss results versus the altitude of UAV. (a) 1 GHz. (b) 4 GHz.

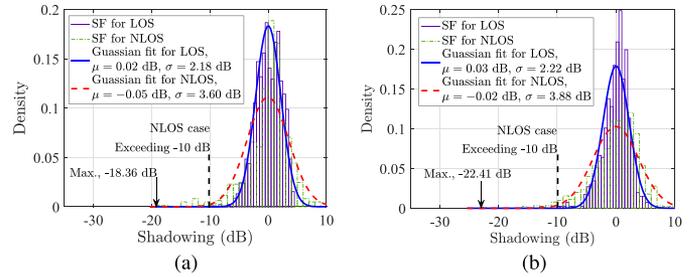

Fig. 5. PDFs of SF for LOS and NLOS condition. (a) 1 GHz. (b) 4 GHz.

where $d$ is the Tx–Rx separation distance in kilometers and $f$ is the carrier frequency in megahertz. In addition, the 3GPP Technical Report 38.901 [8] presents a 3-D path loss model under the LOS and NLOS conditions in the UMa scenario given by

$$PL_{3\text{GPP}} = \begin{cases} 28 + 22\log_{10}(d_{3D}) + 20\log_{10}(f), & \text{LOS} \\ 13.54 + 39.08\log_{10}(d_{3D}) \\ \quad + 20\log_{10}(f) - 0.6(h - 0.5), & \text{NLOS} \end{cases} \quad (2)$$

where $h$ is the altitude of UAV. $d_{3D}$ is in meters, and $f$ is in gigahertz. Note that the model for the LOS case can be used when the horizontal distance $d_{2D}$ is from 10 to $d_{\text{break}} = 4h_{\text{TX}}h_{\text{RX}}/\lambda$, where $\lambda$ is the wavelength.

With comprehensive considerations of the large-scale and small-scale fading, we present an altitude-dependent propagation loss model for the vertical flight of UAV, which can be expressed as

$$PL[\text{dB}] = PL_0[\text{dB}] + X_\sigma + F_\alpha \quad (3)$$

where $PL$ is the propagation loss including the path loss $PL_0$, the shadowing $X_\sigma$, and the fast fading $F_\alpha$. $X_\sigma$ is zero-mean Gaussian random variable occurring at dozens of wavelengths with standard deviation $\sigma$ (also called as shadowing factor) and $F_\alpha$ representing the fast fading (FF) that occurs at the level of wavelength [9].

Based on the models mentioned above, we present the measurement results and conduct comprehensive analysis herein. To reduce the deviation, the data preprocessing was conducted to select a set of data with the most stable flight of UAV from five independent flights, and the received power is averaged by round-trip measurement data. Fig. 4 shows the measurement results, its moving average with 40 wavelengths, the 3GPP models, the free-space model, and the results of our fits at 1 and 4 GHz, respectively. The measurement results in Fig. 4 are obtained from two independent measurements for two frequencies. In order to find the correlation between the path loss and frequency, the altitude of UAV and the horizontal distance between Tx and Rx are the same for each measurement. By referring to the form of the 3GPP model and using the least square (LS) fit, we propose an altitude-dependent path loss model, which is expressed as

$$PL_0 = \begin{cases} 40.55 + 20\log 10(d_{3D}) + 20\log_{10}(f) - n_{\text{LOS}}h \\ 62.41 + 20\log 10(d_{3D}) + 20\log_{10}(f) - n_{\text{NLOS}}h \end{cases} \quad (4)$$

where $n_{\text{LOS}}$ and $n_{\text{NLOS}}$ represent the altitude impact factors for LOS and NLOS case, respectively.

It is evident that as the altitude of UAV rises, the path loss decreases because the channel condition is better in higher altitudes (GS height: 25 m). However, the reduced slope of path loss in NLOS case (below 11 m) is more significant than the LOS case. Moreover, we can see that both the free-space path loss model and the 3GPP model have poor performance in predicting the path loss. However, our model presents a good prediction for the path loss.

Through the LS fits, $n_{\text{LOS}} = 0.102$ and $n_{\text{NLOS}} = 1.190$ at 1 GHz and $n_{\text{LOS}} = 0.250$ and $n_{\text{NLOS}} = 2.075$ at 4 GHz. From the results, it is found that the correlation with height in the NLOS case is stronger than the LOS case, which is consistent with the idea of the 3GPP model. We can see that the small values of $n_{\text{LOS}}$ indicate the weak correlation, which is the reason why the 3GPP model does not introduce the impact factor in the LOS case. Besides, the correlation with height at the higher frequency in the NLOS case is stronger than the lower frequency. Since the impact factors are related to the frequency, it is unreasonable to use a single factor [0.6 in (2)] in the 3GPP model from 0.5 to 100 GHz. However, two frequencies are not enough to find a closed-form expression between the impact factor and frequency. Thus, we only describe the characteristics of correlation. More frequency bands should be investigated in future research.

*B. Shadow Fading*

Although there is no real shadowing in the LOS case, the shadow fading (SF) characterizes the large-scale fluctuation of path loss [9]. We obtain the shadowing data by averaging the instantaneous received power over a 40-wavelength sliding window [10]. Our measurements suggest that the zero-mean Gaussian distribution fits the shadowing under LOS and NLOS conditions in each frequency.

As shown in Fig. 5, the probability distribution functions (PDFs) in each condition are plotted. From the results, it is obvious that the mean values are close to 0, and the shadowing factor $\sigma$ under NLOS is larger than the LOS case. Moreover, there is no significant difference between the two frequencies, which indicates the shadowing is mainly related to the propagation environment. It can be observed that the large shadowing ($\geq 10$ dB) occurs in NLOS condition for both frequencies. The maximum SF can reach 18.36 and 22.41 dB at 1 and 4 GHz, respectively.

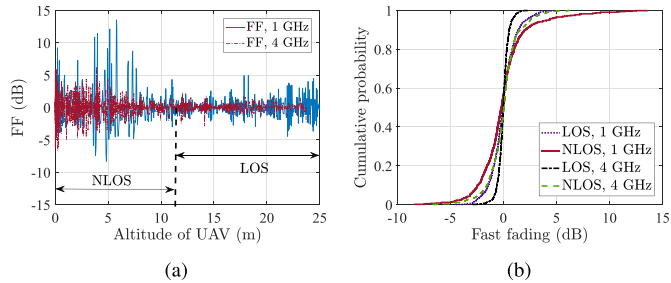

Fig. 6. (a) FF at 1/4 GHz. (b) CDF of FF at 1/4 GHz under LOS and NLOS.

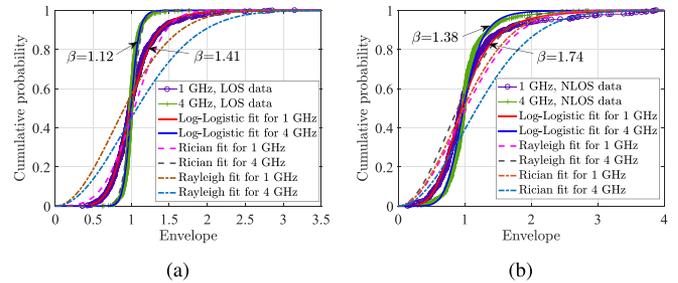

Fig. 7. Statistical distributions of FF. (a) LOS condition. (b) NLOS condition.

TABLE II
RESULTS OF FAST FADING

| $f$ | Link | Max. fading value(dB) | 1%(dB) | 50%(dB) | Fading depth(dB) |
|---|---|---|---|---|---|
| 1 GHz | LOS | 4.93 | -2.87 | -0.052 | 2.818 |
| | NLOS | 13.53 | -5.88 | -0.169 | 5.711 |
| 4 GHz | LOS | -2.88 | -1.62 | 0.009 | 1.629 |
| | NLOS | -6.64 | -4.58 | 0.063 | 4.634 |

## C. Fast Fading

The FF is obtained by averaging the instantaneous received power over a $\lambda/2$-wavelength sliding window [9]. As shown in Fig. 6(a), the FF at 1 GHz is more severe than 4 GHz. For statistical analysis, the cumulative distribution functions (CDFs) of FF under LOS and NLOS conditions in each frequency are drawn in Fig. 6(b).

By counting the CDF data of the FF, the results are listed in Table II. Since most multipath components are with low power, the fades are not severe in the selected scenario. As an example, the maximum fading value and fading depth under NLOS case at 1 GHz is 13.53 dB and 5.711 dB, respectively. Note that the fading depth is defined in terms of the difference in signal levels between the 1% and 50% values [11]. From Table II, we can observe that both of the maximum fading value and the fading depth value present two common phenomena. One is that the channel under NLOS condition experiences larger fading than the LOS case, which is common sense. Another is that the channel in the lower frequency presents more extensive fading behavior than the higher frequency. It is well known that the multipath effect mainly causes the FF. Due to the larger number of multipaths in the lower frequency according to [12], the propagation channels are more natural to experience severe fading. These results are applicable to determine the required fade margin of the UAV-based communication systems.

## D. Statistical Distribution of FF

It is essential to find a match of the fast fading distribution with the theoretical model. In our fitting for CDF of the envelope of FF in the MATLAB, it is found that the log-logistic distribution possesses the maximum value of log likelihood, which represents the log-logistic distribution is the best fit among various distributions including the Nakagami-$m$, Rician, Rayleigh, and Weibull distributions. As shown in Fig. 7, the CDFs of the envelope for LOS and NLOS with the log-logistic, Rician, and Rayleigh fits are plotted. The results show that the fitness of log-logistic is better than the other fits for both conditions.

The PDF of the log-logistic distribution is expressed as

$$f(x,\alpha,\beta) = \frac{(\beta/\alpha)(x/\alpha)^{(\beta-1)}}{(1+(x/\beta)^\beta)^2} \quad (5)$$

where $\alpha$ is a scale parameter and is also the median of the distribution, while $\beta$ is a shape parameter that determines the steepness of the distribution curve. In our fitting, $\alpha$ for all conditions are close to 1, which represents the median value with probability 0.5. It also can be found that the higher $\beta$ leads to the steeper curve, which indicates more severe fading. The values of $\beta$ are 1.41/1.12 and 1.74/1.38 at 1/4 GHz for LOS and NLOS, respectively. The results of $\beta$ show that the channels in the NLOS condition and low frequency are vulnerable to experience more severe fading.

It is also critical to give a physical explanation for the log-logistic distribution. Authors in [13] found that the log-logistic distribution is the best fit for the fading amplitude in foliage environment. Meanwhile in our measurement environment shown in Fig. 1(a), there are also many trees in the environment, which implies that multipaths from trees mainly contribute to the statistical distribution of fading.

## IV. CONCLUSION

In this letter, we have proposed an improved path loss model for NLOS and LOS conditions and comprehensively analyzed the shadowing and fast fading. For the path loss, the 3GPP model only uses a single height impact factor, 0.6 for 0.5–100 GHz in NLOS condition. However, from the measurements, the impact factors are highly related to the frequency. Moreover, the factors under NLOS case in high frequency are larger than that in low frequency. The results of shadowing indicate that the shadowing is related to the environment. Under NLOS case, the shadowing can be larger than 10 dB and reach 18.36 and 22.41 dB at 1 and 4 GHz, respectively. Besides, the fast fading presents strong correlations with the frequency and the link connection. However, the relationship between the fading and frequency shows the negative correlation. The results of our study provide some insights into the fading behavior in AG propagation channels and can be used in designing UAV wireless communications.